\documentclass[letterpaper]{article} 
\usepackage{aaai24}  
\usepackage{times}  
\usepackage{helvet}  
\usepackage{courier}  
\usepackage[hyphens]{url}  
\usepackage{graphicx} 
\usepackage{natbib}  
\usepackage{caption} 
\usepackage{booktabs}
\usepackage{multirow}
\usepackage{filecontents}
\usepackage{adjustbox}
\usepackage{pgfcalendar}

\usepackage{algorithm}

\usepackage{newfloat}
\usepackage{listings}
\usepackage{pgfplots}
\usepgfplotslibrary{dateplot}
\usepackage{adjustbox}

\usepackage{pgfplotstable} 

\DeclareCaptionStyle{ruled}{labelfont=normalfont,labelsep=colon,strut=off} 
\floatstyle{ruled}
\newfloat{listing}{tb}{lst}{}
\floatname{listing}{Listing}
\title{``Every Corner is a Crime Scene'': Capturing Narrative Evolution Among Ukrainian and Russian Communities in the Russo-Ukrainian War}
\title{``A Provocation Committed by [Russia/Ukraine]'': Capturing Narrative Evolution Among Ukrainian and Russian Communities in the Russo-Ukrainian War}
\title{``We Will Show you Real Denazification'': Capturing Narrative Evolution in Both Sides of the Russo-Ukrainian War}
\title{Modeling Information Narrative Detection and Evolution on Telegram\\ during the Russia-Ukraine War}

\author{
    Patrick Gerard\textsuperscript{\rm 1,\rm 2}, 
    Svitlana Volkova\textsuperscript{\rm 2}, 
    Louis Penafiel\textsuperscript{\rm 2}, 
    Kristina Lerman\textsuperscript{\rm 1}, 
    Tim Weninger\textsuperscript{\rm 3}
}
\affiliations{
    \textsuperscript{\rm 1}Information Sciences Institute, University of Southern California, Marina del Rey, California, United States\\
    \textsuperscript{\rm 2}Aptima, Inc., Dayton, Ohio, United States\\
    \textsuperscript{\rm 3}University of Notre Dame, Notre Dame, Indiana, United States\\
    pgerard@isi.edu, pgerard@aptima.com, svolkova@aptima.com, lpenafiel@aptima.com, lerman@isi.edu, tweninger@nd.edu
}

\usepackage{bibentry}
\usepackage{algpseudocode}

\begin{document}

\maketitle

\begin{abstract}
Following the Russian Federation's full-scale invasion of Ukraine in February 2022, a multitude of information narratives emerged within both pro-Russian and pro-Ukrainian communities online. As the conflict progresses, so too do the information narratives, constantly adapting and influencing local and global community perceptions and attitudes. This dynamic nature of the evolving information environment (IE)  underscores a critical need to fully  discern how narratives evolve and affect online communities. Existing research, however, often fails to capture information narrative evolution, overlooking both the fluid nature of narratives and the internal mechanisms that drive their evolution. Recognizing this, we introduce a novel approach designed to both model narrative evolution and uncover the underlying mechanisms driving them. In this work we perform a comparative discourse analysis across communities on Telegram covering the initial three months following the invasion. First, we uncover substantial disparities in narratives and perceptions between pro-Russian and pro-Ukrainian  communities. Then, we probe deeper into prevalent narratives of each group, identifying key themes and examining the underlying mechanisms fueling their evolution. Finally, we explore influences and factors that may shape the development and spread of narratives.\footnote{Code to reproduce our models and findings available at \url{https://anonymous.4open.science/r/Uncovering-Narratives-E433/}}
\end{abstract}
\section{Introduction}
Operations within the Information Environment (OIE) rely crucially on the strategic deployment of information narratives, as outlined in JP3-04, which describes them as \textit{a method to convey or interpret events in a manner that supports a specific perspective or set of values} (ibid. p. I-7). These narratives play a critical role in shaping how audiences perceive events, a challenge that has intensified with the advent of digital dissemination tools like generative AI \cite{goldstein2023generative} and social media platforms. The increasing accessibility and influence of these tools underscore the urgent need for advanced methods capable of effectively tracking and analyzing narratives. Such analysis is vital for unraveling the complexities of how narratives are crafted, evolve, and exert influence across the information landscape.

The ongoing war between Russia and Ukraine epitomizes the critical role of such narratives, with varying claims—from Russia's purported mission to "denazify" Ukraine to allegations of biological weapons labs—significantly influencing public opinion~\cite{volkova2021machine}. Conventional research methodologies often fall short in capturing the fluid dynamics of narratives, which adapt and transform in response to emerging information and evolving contexts~\cite{volkova2024}.

To bridge this methodological divide, our study proposes an innovative approach tailored to the dynamic analysis of information narratives. It leverages a two-tiered framework designed to not only identify but also analyze the complexity of narrative constructs across social platforms. This method allows for a comprehensive analysis, from a broad overview of narrative structures to an in-depth examination of individual components, shedding light on narrative evolution—how they emerge, fade, and transform over time.

Applying this methodology, we delve into the narratives propagated within Russian and Ukrainian Telegram communities during the initial phase of the war. Our analysis illuminates stark disparities in narrative engagement and perception between these communities. We explore the predominant themes and the underlying dynamics driving their evolution within each community. Additionally, we investigate the key figures and forces shaping the dissemination and evolution of these narratives, offering a nuanced understanding of their influence and trajectory.
\section{Related Work and Background}
The exploration of narrative evolution within online communities, particularly in response to geopolitical events, has been addressed through various methodologies across the fields of computational social science. Prominent among these methodologies are clustering algorithms and topic modeling techniques, each offering unique insights but also exhibiting notable limitations in the context of dynamic narrative analysis.
\vspace{-1mm} 
\subsection{Narrative Detection}
The concept of 'narrative' has been a longstanding subject of study within the discipline of narratology \cite{smith1980narrative,labov2013language}; however, while the field generally agrees on the components of narratives, it still lacks a standard definition. Previous studies posit ``narrative frameworks'' as being comprised of stories which may align otherwise disparate domains of knowledge \cite{corona-conspiracy}. To build on this definition and follow the JP3-04, we adapt an information narrative framework detailed and used by similar recent works on narratives \cite{hanley2023specious}, which utilize the logic of the Event Registry \cite{event-registry} idea where collections of documents seek to address the same event or issue \cite{miranda-etal-2018-multilingual}. 
Applying this logic, a narrative in our dataset might be ``There are Biological Weapons Laboratories in Ukraine." This narrative encompasses various stories, each consisting of individual posts. For example, the text ``We confirm the facts that were revealed of the emergency cleansing by the Kiev regime of traces of the military-biological program" would be part of the ``Cover-up and Destruction of Biological Weapon Development" story, which in turn falls under the broader narrative about biological weapons laboratories in Ukraine.
\vspace{-1mm} 
\subsection{Information Spread and Narrative Evolution}
The dynamic nature of online information dissemination underscores the critical need for advanced analytical tools to capture the fluid evolution of narratives. Computational narrative analysis has emerged as a key approach, providing methodologies for dissecting and understanding the complex interplay between narrative structures and their propagation through digital platforms.

Significant contributions, such as Chambers and Jurafsky's work on narrative event chains, and McIntyre and Lapata's developments in narrative generation and machine learning, have established a strong foundation for understanding narrative dynamics \cite{chambers-jurafsky-2008-unsupervised, mcintyre-lapata-2009-learning}. However, the rapid spread of online narratives often challenges traditional methods, leaving some complex aspects and internal dynamics of narrative evolution unaddressed.
\vspace{-1mm} 
\subsection{Narrative Dynamics during  RU-UA War}
These challenges becomes particularly evident in the context of the Russian-Ukrainian war. The strategic deployment of narratives, from claims of "denazification" to allegations of biological weapons labs, has significantly influenced public perception, both locally and globally \cite{jowett2018propaganda, badawy2019characterizing}. The marked increase in narrative dissemination through various media channels during this conflict highlights the need for more dynamic and adaptable tools in narrative analysis, capable of tracking the evolution of such narratives in real-time.
\vspace{-2mm} 
\subsection{Clustering and Topic Modeling Limitations}While techniques like K-means, hierarchical clustering, Latent Dirichlet Allocation (LDA), and BERTopic have advanced text grouping and theme identification, they struggle in the dynamic field of narrative analysis due to their need for predetermined cluster numbers \cite{blei2003latent, grootendorst2020bertopic}. This requirement limits flexibility and fails to adapt to the unpredictable emergence of narratives in response to real-world events \cite{aggarwal2012survey, jain2010data, mcinnes2017hdbscan}.

Additionally, these methods offer only static snapshots, inadequate for monitoring evolving narratives that change with new information and shifts in discourse. Although DBSCAN provides an alternative by not requiring preset cluster numbers, adapting it for real-time, streaming data environments remains a significant challenge. Its focus on data density may also misrepresent the complex structure of narratives, hindering effective analysis in digital contexts.

To address these issues, we propose a novel framework for narrative detection and evolution that adjusts to new information, identifies emerging narratives, and integrates domain expertise for a deeper, more nuanced analysis.

\section{Methodology}
This section outlines our method for dynamically clustering text data streams over time, designed to adapt to new content. It aims to monitor the evolution of text clusters, aiding in identifying how narratives diverge or converge over time. This approach supports the organic discovery of narrative structures as they form, making it well-suited for analyzing the fluid information streams . An additional analytical layer can further enhance this model by integrating expert interpretation with clustering results, providing deeper narrative analysis. This component, while optional, merges stories into cohesive narratives, enriching the algorithmic insights. While the primary model effectively manages temporal clustering, this integrated methodology is particularly aimed at revealing the complex dynamics within narratives.

\vspace{-2mm} 
\subsection{Our Model -- Clustering Streaming Text Data}
The dynamic nature of analyzing evolving narratives demands a model that not only allows for the inherent interpretability of its processes but also facilitates the dynamic discovery of narrative structures. Hierarchical Agglomerative Clustering stands out as our method of choice, given its unique capacity to meet these requirements effectively.

\vspace{-2mm} 
\paragraph{Hierarchical Agglomerative Clustering (HAC) for Dynamic Narrative Analysis} HAC is recognized for its adaptability in data analysis, particularly in forming clusters based on similarity without the need for pre-defined cluster numbers. This approach begins with each data point as a separate cluster, iteratively merging the most similar pairs until a specified similarity threshold, such as cosine similarity, is met, ensuring homogeneity within clusters \cite{mullner2011modern}. Its inherent ability to uncover data structures naturally positions HAC as a prime clustering candidate for extension for fluid narrative analysis, where the evolving nature of text-based data necessitates flexible and dynamic clustering techniques. Furthermore, HAC's iterative merging process is apt for online extensions, allowing new data to be integrated efficiently in real-time. This adaptability ensures ongoing narrative analysis remains current, making HAC an exemplary foundation for methodologies aimed at exploring narrative evolution in dynamic environments.

\vspace{-2mm} 
\paragraph{Adapting HAC to Evolving Data Streams}
To adapt HAC for real-time data analysis, we developed the \texttt{OnlineAgglomerative} class, which enhances HAC for dynamic environments. This class is detailed in our GitHub repository.$^1$ 
Key to this adaptation is the \texttt{incremental\_fit()} method, which dynamically clusters data and adjusts to new inputs by balancing immediate data integration with long-term cluster evaluation. Initially, clusters are formed using HAC. The \texttt{incremental\_fit()} method then uses the class's cluster history to decide at each timestep whether to merge new data into existing clusters or initiate new clustering rounds, based on a predefined semantic similarity threshold.

Additionally, \texttt{incremental\_fit()} evaluates potential cluster mergers with each data batch using Silhouette scores, which measure cluster cohesion and separation, and the pseudo-F index, assessing cluster quality. This approach not only ensures immediate data integration but also maintains the structure and coherence of clusters over time. As a result, our approach  enables effective real-time clustering by making immediate decisions and maintaining complex data structures as new information arrives, thereby helping to discover underlying data patterns in dynamic settings. 
 
\paragraph{Exploiting Online Data Integration} 
Building on this foundation, the \texttt{OnlineAgglomerative} class then enhances our ability to exploit streaming data integration effectively. Through the \texttt{incremental\_fit()} method, it tracks each cluster's centroid and size history, offering insights into cluster evolution. This tracking reveals patterns of development, merging, or separation, allowing us to decode the evolving landscape of data and offering a window into the complex mechanisms that drive the data's evolution.

\vspace{-2mm} 
\subsection{Advancing Our Model -- Macro-Narratives}
To enhance our model, we then introduce the \texttt{MacroNarrative} Class, an additional layer atop the \texttt{OnlineAgglomerative} curated clusters. This optional augmentation allows domain experts to weave together these clusters into a coherent narrative framework. We find this to be especially vital given the inherently fragmented landscape of narrative frameworks (further demonstrated in our data analysis), allowing for a synthesized, expert-informed view of the narrative's progression and thematic undercurrents.

\vspace{-2mm} 
\subsection{Macro-Narrative Detection}
After tracking story clusters using the \texttt{OnlineAgglomerative} class, we then use these as ``story clusters'' and employ the \texttt{MacroNarrative} class to infuse expert knowledge and examine how these story clusters form larger narrative clusters. 

\vspace{-2mm} 
\paragraph{Step 1: Discovery and Refinement of Narrative Seeds}


To begin our formation of narrative clusters, we start with a single story cluster uncovered in the previous step. This story cluster serves as the initial `seed' from which we methodically unravel additional seeds, laying the groundwork for constructing a comprehensive narrative cluster.  

\vspace{-2mm} 
\paragraph{Step 2: Human-in-the-Loop (HITL) Refinement}
Upon identifying an initial potential seed cluster, our next step involves seeking additional seed clusters to guide the development and progression of a \textit{Narrative Cluster}. This procedure harmonizes automated discovery with HITL refinement, engaging domain experts to ensure a balanced approach. Initially, we employ a queue-based system for temporal evaluation of the seed. At each timestep, story clusters falling within a predefined similarity threshold—typically between 0.7 and 0.8 cosine similarity, as determined effective for our narratives—are queued for further analysis and simultaneously earmarked for human review. This method, reminiscent of a breadth-first search (BFS), ensures a comprehensive and \textit{robust} selection of potential seeds for additional refinement by analysts. These analysts then evaluate a representative sample, chosen randomly (in our scenario, 20 points from each cluster), at every timestep. The analysts then save clusters that retain ongoing \textit{relevance} and \textit{alignment} with the intended narrative direction. These are used as the ``seed clusters'' whose positions in the embedding space will guide the resulting Narrative Cluster throughout time.

\vspace{-2mm} 
\paragraph{Step 3: Narrative Centroid Construction}
Expanding from the curated seed clusters, the \texttt{MacroCluster} class constructs a 'Narrative Centroid' that dynamically navigates through time alongside its seed clusters. This evolving centroid adjusts in real-time to the development of seed story clusters, enabling the narrative cluster to reflect the narrative's fluidity over time. By applying similarity thresholds, the \texttt{MacroCluster} class then integrates other story clusters into the narrative at various points in time, offering an adaptive perspective on narrative evolution.

An example of this adaptive narrative tracking is our analysis of information narratives during the Bucha Massacre in 2022. In examining the Bucha Massacre narratives within the Ukrainian community, we observed the interplay of various story clusters pertaining to the liberation of Bucha within the narrative cluster. While these clusters were initially identified as during the automated discovery stage, their generalized focus on liberation precluded them from being considered as grounding seeds. However, during the appropriate time period (i.e., when they were discussing the liberation of Bucha), their interactions with the established narrative were still captured and analyzed through the narrative centroid's adaptive framework, demonstrating the ability to reflect both the stability and fluidity of narrative elements within a broader context.

\vspace{-1mm} 
\section{Data Analysis: Russo-Ukrainian War}



Applying our \texttt{OnlineAgglomerative} and \texttt{MacroNarrative} classes, we then delve into the narratives emerging from the Russo-Ukrainian War. This conflict, known for its complex and evolving narratives, provides a fertile testing ground for our methodology. It challenges our model to adeptly cluster and track these shifting narratives, offering a robust validation of our approach in a context filled with a rich tapestry of political and human interest stories.



\subsection{Data Collection}
We use posts collected from Russian-oriented and Ukrainian-oriented Telegram channels \cite{theisen2022motif} spanning from October 2022 to August 2023. These channels and their posts were collected using a combination of an expert-generated queue of telegram channels and snowballing via those channels to perform segmentation analysis described below. 

Overall, there are 989 channels represented with over 9.67 million total posts, written mostly in Ukrainian and Russian. We sampled this data to focus our analysis on the first three months of the war (from February 20, 2022 through May 28, 2022), in which we track 568 channels (comprising distinct ``Russian-leaning'' and ``Ukrainian-leaning'' communities) and approximately 2 million posts.

\paragraph{Telegram Platform}
Telegram, a messaging app that facilitates both user interactions in private and public groups and one-way broadcasts via channels, is seen as a bastion of a "free" Internet in Russia, evading bans that affect other platforms like Facebook and TikTok since 2020 \cite{oleinik2024telegram}. It has emerged as a vital platform for military bloggers and a primary information source on the Russo-Ukrainian war, with approximately 39\% of Ukrainians and 19\% of Russians relying on it for news, ranking it highly in information sourcing in both countries \cite{oleinik2024telegram}. This positions Telegram as a crucial subject for analyzing discourse and narrative evolution regarding the conflict.

\definecolor{darkspringgreen}{rgb}{0.09, 0.45, 0.27}

\paragraph{Network Construction and Data Partitioning}

Given that our data was collected to focus on the Russia-Ukraine war, it consists of mostly Ukrainian-centric and Russian-centric communities. To understand how each community discusses and understands the war in the beginning of the invasion, we  first separate the data into these communities. To do this, we construct author networks amongst Telegram users and then use label-propagation algorithm to separate these networks \cite{garza2019community}.


\paragraph{Community Discovery}
We aim to discover relatively homogeneous communities (Ukrainian-centric or Russian-centric) within our data so that we may understand and compare how each community discusses the conflict. To construct our community networks, we lean on  previous findings that retweets act as a relative indicator of endorsement-based connections \cite{metaxas2015retweets}. For Telegram data, we build a reference network of channels where a directed link with weight \textit{w} connects channel \textit{A} to channel \textit{B} if \textit{A} references or forwards a post of \textit{B} \textit{w} times within the period. 





\paragraph{Community Network Partitioning} For Telegram channels, we are able to view the biography and recent posts of each channel using telegrams native channel search service (t.me/channel\_name). This allows us to inspect random ``seed channels'' that can guide the propagation of labels. Utilizing 3 classes ``Ukrainian-centric'', ``Russian-centric'', and ``Other'', we inspect and label 100 random ``seed'' channels (approximately 14\% of channels) and run the label-propagation algorithm on the Telegram network.

\paragraph{Data Sample Statistics and Validation} We inspect 75 random channels from each partition to ensure the validity of each partition. Overall, there are 243 Ukrainian-leaning channels with a median of 71K posts per timestep (and a total of 4.2 million posts) and 325 Russian-leaning channels with a median of 77K posts per timestep (and a total of 4.4 million posts).




\subsection{Data Preprocessing}
For each post, we first remove all URLs, emojis, and  hashtags. Next, we remove any duplicate posts. We define a duplicate post as having the same text and author so as to avoid duplicates that occur likely by mistake while preserving duplicates that arise out of the observed accounts copying each other's posts. Next, following the findings of previous work \cite{hanley2023partial}, we remove any posts that have fewer than four words. Finally, we break up the post into 2-sentence texts; this follows the logic and findings of prior works which posit that posts or articles often address multiple narratives but that smaller sentence-level components will typically discuss the same
narrative \cite{piktus2022web, hanley2023partial}.

\paragraph{Text Embedding Encoding}
We utilize the multilingual MPNet embedding model \cite{song2020mpnet} to embed each text; specifically, we use the MPNet model fine-tuned for clustering and semantic search\footnote{\url{https://huggingface.co/sentence-transformers/paraphrase-multilingual-mpnet-base-v2}}. We chose it due to its ability to handle 50 languages (including the primary languages in our dataset: Russian, Ukrainian, and English) as well as its performance on similar tasks \cite{hanley2023partial, hanley2023specious}.

\paragraph{Semantic Similarity Measures}
We utilize cosine distance (rather than Euclidean or Manhattan distance, for example) to drive our clustering algorithm due to cosine similarity's (where cosine distance = 1 - cosine similarity) observed relationship with semantic similarity \cite{rahutomo2012semantic}. To guarantee high semantic similarity in our story clusters, we tested various cosine similarity thresholds, ranging from 0.60 to 0.85 in increments of 0.05, for grouping  topically texts. Following prior research suggesting a 0.60-0.80 range for topical similarity \cite{hanley2023happenstance}, we had a researcher label 200 randomly paired texts at each threshold as either ``semantically similar'' or ``not semantically similar.'' We then compared these labels to each threshold's predictive accuracy of semantic similarity. Our findings indicate that a threshold of 0.85 yields the highest accuracy, which is in line with similar work \cite{hanley2023happenstance}.

\subsection{Data Analysis and Validation}
Following our data preprocessing, we then employ our \texttt{OnlineAgglomerative} and \texttt{MacroCluster} to uncover and dissect narratives from the  Russian and Ukrainian online communities on Telegram. First, we segregate stories using the \texttt{OnlineAgglomerative} class for clustering. Then, infusing domain expert insights, we form narrative clusters using the \texttt{MacroCluster} class, capturing both predetermined and newly identified narratives through trend analysis. Finally, we distill key themes from these narratives and analyze their progression over time.

\paragraph{Story Cluster Discovery \& Validation} 
We automatically dynamically cluster the data from either community using the \texttt{OnlineAgglomerative} class's \texttt{incremental\_fit()} method.

To ensure the validity of the evolving clusters, we must monitor that each cluster (1) is \textit{cohesive} within a timestep and these results are (2) \textit{consistent} across timesteps. To do this, we ran our incremental fitting process across 15 weeks, each week comprising an average of 120 thousand messages. We find that each point had an average cosine similarity of 0.935 with its respective cluster centroid on any given timestep -- which is above our set similarity threshold and indicating \textit{cohesion} -- and an average cosine similarity of 0.245 with remaining clusters. Furthermore, we find that these similarity results are \textit{consistent} across timesteps (i.e., it does not degrade as our algorithm considers more timesteps), suggesting that the algorithm effectively encompasses new data, correctly discerning when to fold data into existing clusters (and adjust the characteristics of these clusters) or create new clusters.

\paragraph{Trending Story Discovery} 
Following the \texttt{OnlineAgglomerative} clustering, our methodology includes a key process for identifying emerging trends within story clusters, utilizing the \texttt{analyze\_micro\_cluster\_trends()} function, included in our code repository. This approach  evaluates the expansion or reduction of story clusters at each timestep by tracking the change in data volume compared to the preceding period. More than just pinpointing the largest clusters at any given time (which often reflected persistent themes rather than trending topics) our strategy focuses on identifying fluctuations that signal shifting interests or concerns within the community; this method proved particularly effective in highlighting stories gaining traction and capturing the community's engagement with unfolding events. These insights revealed not only what topics are currently engaging the community but also serve as indicators for potential overarching narratives, laying the groundwork for deeper narrative analysis.

\paragraph{Trending Story Validation} 
Our analysis focused on the 5 top-trending clusters each week, revealing that approximately 93.2\% of the time, these clusters directly matched significant, unfolding external events documented in the accompanying documentation with this paper to ensure reproducibility. This underscores our model's proficiency in identifying relevant narrative stories. Notably, our dynamic methodology—leveraging information from the previous timestep to inform the current one—demonstrated superior performance compared to static approaches that correlate trendiness solely with cluster size within the same timestep. Detailed findings are further elaborated in the provided documents in the repository.

\paragraph{Narrative Cluster Formation and Validation}
Next, collaborating with a domain expert, a former analyst, we formed narrative clusters using the \texttt{MacroCluster} framework for both Pro-Russian and Pro-Ukrainian communities. These narratives are shown in tables \ref{tab:uk_narratives} and \ref{tab:ru_narratives}. We note that while most of the narratives we sought to study were predetermined (for example, "Russian troops committed a massacre in Bucha"), others were only discovered by evaluating the trending story clusters (for example, "Russia is engaging in chemical warfare" and "Russia is sabotaging Ukrainian humanitarian corridors"). 

To validate our narrative clusters, we cross-referenced the emerging and predetermined narratives against external sources and expert opinions. This process ensured that the clusters accurately reflected the evolving discourse within the targeted communities.

\paragraph{Additional LLM-Driven Validation}
Following our story and narrative clustering and monitoring of narrative macro-clusters, we sought to understand what themes emerge and how themes evolve within narratives. To do this, we employed a combination of autoregressive and autoencoding models - LLama2 Large Language Model (LLM) \cite{touvron2023llama} and a multilingual Deberta model \cite{he2023debertav3} fine-tuned using zero-shot classification\footnote{ https://huggingface.co/MoritzLaurer/mDeBERTa-v3-base-xnli-multilingual-nli-2mil7} to perform  narrative theme extraction and classification. 
%


\paragraph{Narrative Theme Extraction}
We employ off-the-shelf model described above, which has not seen the specific categories during training to classify texts from narrative clusters based on context and semantics \cite{wang2019survey}. We use a multi-label setting because we found that some texts can indeed represent more than one theme effectively. Then the model then returns a list of corresponding model scores and confidences for each label.


\paragraph{Narrative Theme Extraction Validation}
We aim to distill narrative themes that provide both broad \textit{coverage} and detailed \textit{nuances}. The process begins by generating 15 theme dictionaries through analyzing story clusters within the similarity thresholds of narrative centroids. We initialize an empty map to track the emergence of each theme. The Llama2 model, supplied with data points closest to each cluster's centroid and any previously generated themes, identifies new themes or refines existing ones. Each theme and its emergence timestep are recorded in this map. The process is thoroughly documented in our repository.

We then evaluate each theme dictionary using a classification model to calculate a Theme Coverage Score (TCS), which measures the percentage of texts exceeding a set threshold for at least one theme, summed across all timesteps. A higher TCS indicates better theme recognition, suggesting a more comprehensive and representative set of themes. After comparing TCS values, we select the dictionary with the highest score, ensuring our narrative analysis is accurate and representative. While TCS scores are generally consistent, variations occur mainly in the number of themes and their phrasing. The validation of the classification model’s scoring mechanism was conducted both prior to and during the theme extraction processes.
\paragraph{Narrative Theme Classification}
In the final step, we use the final collection of extracted themes and applied the classification model to assign thematic labels to each message within the narrative. For a message to be classified under a specific theme, it had to meet or exceed a predetermined confidence threshold (outlined in the subsequent paragraph), ensuring that our thematic categorization was both precise and meaningful.

\paragraph{Narrative Theme Classification Validation}
To validate theme classification, we establish a ``confidence threshold'' for each narrative. Text must exceed this threshold (0-1) to be assigned a specific theme. We determine this threshold by initially running our theme extraction model, then manually examining how the classification model labels 200 texts at varying thresholds. Prioritizing specificity to minimize noise, we choose a threshold that optimizes accuracy. We note that theme trends stay consistent across various thresholds, reinforcing our confidence in our method's reliability.

\vspace{-1mm} 
\section{Results and Key Findings}
We apply and validate our novel narrative detection and evolution pipeline to the Russian and Ukrainian communities. Then we first evaluate both communities separately, inspecting them at the macro (narrative) level and the micro (story) level before then performing a contrastive analysis during the Bucha Massacre in Ukraine  as an example.

\vspace{-1mm} 
\subsection{Story Cluster Analysis}
To understand how both communities discuss key events during the war, we extract and analyze trending story clusters at each timestep. Our pipeline first outputs translations of 10 messages near each cluster's centroid and 5 random messages, followed by a LLama2-generated summaries. As indicated earlier, we release these summaries, along with news articles related to each summary in our repository. 

\paragraph{Both Communities React Quickly}
Both communities react very quickly and consistently to events. As illustrated in the google sheets document linked in our GitHub repository, in each timeframe examined, the majority of the leading story clusters were closely linked to external events. 

\paragraph{Contrasting Community Reactions to External Events} The events and stories capturing each community's attention differ greatly, reflecting their unique focuses and concerns. Ukraine's trending clusters primarily highlight critical developments like airstrikes and humanitarian efforts, whereas Russia's narratives, often disconnected from ground realities, emphasize narratives like international reactions (frequently framing Russia as a victim) and economic impacts. To get a sense of this disparity, we have a human annotator label the top 10 trending stories from either community for each week as ``Military/War,'' ``Politics - Internal,'' or ``Politics - International;'' the results are shown in Table \ref{tab:trending_stories_makeup}. 

\begin{table}[h]
\centering
\small
\caption{Micro-narrative story clusters on Telegram.}
\label{tab:trending_stories_makeup}
\begin{tabular}{lccc}
\toprule
Community & Military/War & \multicolumn{2}{c}{Politics} \\
\cmidrule(lr){3-4}
 &  & Internal & International \\ 
\midrule
Ukrainian & \textbf{46.5\%} & 9.3\% & 44.2\% \\
\addlinespace
Russian & 16.7\% & 31\% & \textbf{52.4\%} \\
\bottomrule
\end{tabular}
\end{table}
\vspace{-2mm} 

 Table \ref{tab:trending_stories_makeup} reveals significant differences in the primary concerns of each community, particularly within the realm of international politics. Ukrainian narratives often emphasize foreign aid and movements towards EU and NATO integration, while Russian narratives typically highlight economic strategies and portray Russia as subjected to biased international critique.

For instance, in the week following March 6, 2022, after a suspected Russian airstrike intended to draw Belarus\footnote{https://www.reuters.com/world/ukraine-says-russian-aircraft-fired-belarus-ukrainian-air-space-2022-03-11/} into the conflict, Ukrainian discourse focused on the event and its implications, contrasting sharply with Russian discourse which largely overlooked the airstrike. Instead, Russia emphasized allegations of US-backed biological warfare in Ukraine and the exit of Western businesses from Russia, illustrating distinct narrative priorities.

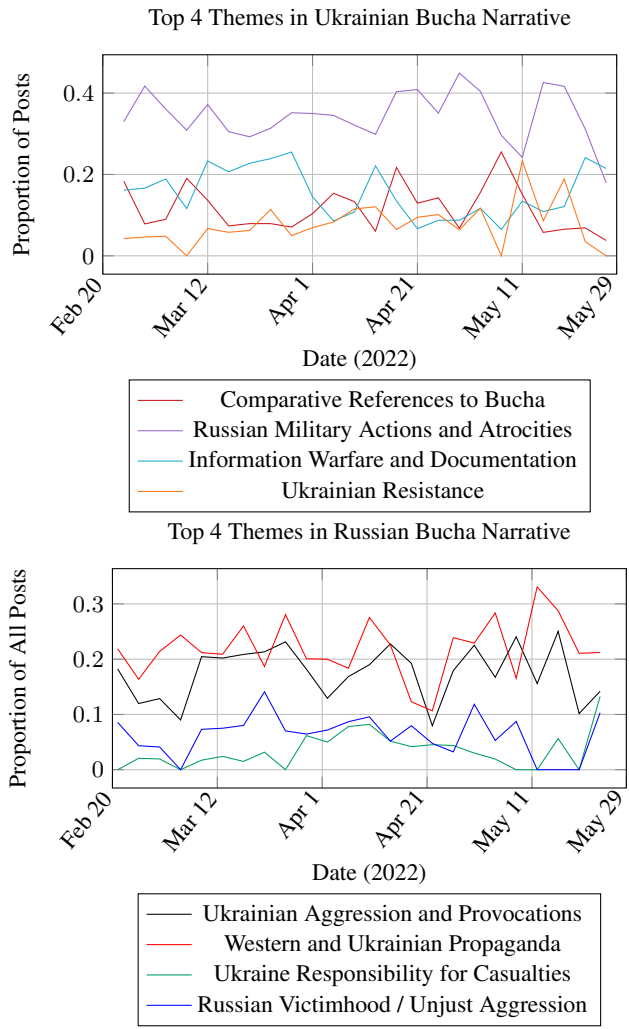
\begin{figure}[ht]
\setlength{\belowcaptionskip}{1pt} 
\centering
\centering
\small
\begin{tikzpicture}
\begin{axis}[
    xlabel={Date},
    legend style={at={(0.5,-.48)},anchor=north},
    ylabel={Score},
    title={Scores Over Time for Different Categories},
    date coordinates in=x, 
    xticklabel={\year-\month-\day}, 
    xticklabel style={rotate=50, anchor=east}, %
    xlabel style={
        yshift=-10.5, 
    },
    ylabel style={
        yshift=-5, 
    },
    ylabel=Proportion of Posts,
    title=Top 4 Themes in Ukrainian Bucha Narrative,
    xlabel=Date (2022),
    grid=both, 
    width=\linewidth, 
    height=4.5cm, 
    xtick={2022-02-20, 2022-03-12, 2022-04-01, 2022-04-21, 2022-05-11, 2022-05-29},
    xticklabels={Feb 20, Mar 12, Apr 1, Apr 21, May 11, May 29},
    xmax=2022-05-29, %
    xmin=2022-02-20, %
    ]

\definecolor{amethyst}{rgb}{0.6, 0.4, 0.8}
\definecolor{ballblue}{rgb}{0.13, 0.67, 0.8}
\definecolor{darkorange}{rgb}{1.0, 0.55, 0.0}
\definecolor{emerald}{rgb}{0.31, 0.78, 0.47}
\definecolor{fireenginered}{rgb}{0.81, 0.09, 0.13}
\definecolor{pumpkin}{rgb}{1.0, 0.46, 0.09}

\addplot+[
  mark=none,
  mark options={solid},
  color=fireenginered
  ]
  table [x=date, y=Comparative References to Bucha, col sep=comma] {LaTeX/uk_bucha_rounded_5.csv};
\addlegendentry{Comparative References to Bucha}

\addplot+[
  mark=none,
  mark options={solid},
  color=amethyst
  ]
  table [x=date, y=Russian Military Actions and Atrocities, col sep=comma] {LaTeX/uk_bucha_rounded_5.csv};
\addlegendentry{Russian Military Actions and Atrocities}

\addplot+[
  mark=none,
  mark options={solid},
  color=ballblue
  ]
  table [x=date, y=Ukrainian Resistance and Military Operations, col sep=comma] {LaTeX/uk_bucha_rounded_5.csv};
\addlegendentry{Information Warfare and Documentation}

\addplot+[
  mark=none,
  mark options={solid},
  color=pumpkin
  ]
  table [x=date, y=Information and Documentation, col sep=comma] {LaTeX/uk_bucha_rounded_5.csv};
\addlegendentry{Ukrainian Resistance}

\end{axis}

\end{tikzpicture}

\label{fig:uk_bucha}
\small
\centering

\begin{tikzpicture}
\begin{axis}[
    xlabel={Date},
    legend style={at={(0.5,-.48)},anchor=north},
    ylabel={Score},
    title={Scores Over Time for Different Categories},
    date coordinates in=x, 
    xticklabel={\year-\month-\day}, 
    xticklabel style={rotate=50, anchor=east}, %
    xlabel style={
        yshift=-10.5pt, 
    },
    ylabel=Proportion of All Posts,
    title=Top 4 Themes in Russian Bucha Narrative,
    xlabel=Date (2022),
    grid=both, 
    width=\linewidth, 
    height=4.5cm,
    xtick={2022-02-20, 2022-03-12, 2022-04-01, 2022-04-21, 2022-05-11, 2022-05-29},
    xticklabels={Feb 20, Mar 12, Apr 1, Apr 21, May 11, May 29},
    xmax=2022-05-29, %
    xmin=2022-02-20, %
    ]

\addplot+[
  mark=none,
  mark options={solid},
  color=black
  ]
  table [x=date, y=Claims of Ukrainian Aggression and Provocations, col sep=comma] {LaTeX/ru_bucha_plottable.csv};
\addlegendentry{Ukrainian Aggression and Provocations}

\addplot+[
  mark=none,
  mark options={solid},
  color=red
  ]
  table [x=date, y=Allegations of Western and Ukrainian Propaganda, col sep=comma] {LaTeX/ru_bucha_plottable.csv};
\addlegendentry{Western and Ukrainian Propaganda}
\definecolor{screamin_green}{rgb}{0.46, 1.0, 0.44}
\definecolor{seagreen}{rgb}{0.18, 0.55, 0.34}
\definecolor{shamrockgreen}{rgb}{0.0, 0.62, 0.38}
\addplot+[
  mark=none,
  mark options={solid},
  color=shamrockgreen
  ]
  table [x=date, y=Claims of Ukrainian Responsibility for Civilian Casualties, col sep=comma] {LaTeX/ru_bucha_plottable.csv};
\addlegendentry{Ukraine Responsibility for Casualties}

\addplot+[
  mark=none,
  mark options={solid},
  color=blue
  ]
  table [x=date, y=Narratives of Russian Victimhood and Unjust Aggression, col sep=comma] {LaTeX/ru_bucha_plottable.csv};
\addlegendentry{Russian Victimhood / Unjust Aggression}




\end{axis}

\end{tikzpicture}

\label{fig:ru_bucha}
\caption{Contrasting information narratives surrounding the Bucha Massacre across Russian and Ukranian communities.}
\label{fig:bucha_narratives}

\vspace{-0.3cm}
\vspace{-1mm} 
\end{figure}

Moreover, when addressing shared topics, the communities exhibit markedly different viewpoints. The Bucha Massacre's coverage showcases this divergence; Ukrainian discussions revolved around exposing Russian crimes and debunking misinformation, whereas Russian discourse framed the coverage as biased, attributing it to Ukrainian provocation. This stark contrast in perspectives, especially evident in the Bucha case, underscores the ongoing narrative division, as depicted in Figure \ref{fig:bucha_narratives}.

\vspace{-1mm}
\subsection{Narrative Analysis}
We conduct an in-depth analysis of four major narratives from each community (shown in Tables \ref{tab:uk_narratives} and \ref{tab:ru_narratives})  evaluating the narratives at both the story and the theme levels; we examine their persistence, the micro-narratives they encompass, their summaries, and the key authors involved. 

\paragraph{Narrative Stories Vary across Communities}
First, we find that the number of story clusters comprising each narrative is narrative-specific. For example, the Russian narrative about denazifying Ukraine contained many more story (micro-narrative) clusters (7-10 depending on the timestep) than the Ukrainian narrative about Ukraine belonging in the European Union (4-7 depending on the timestep). 
Additionally, we find that the number of comprising stories can shift across timesteps and in response to external events. For example, following Russian accusations of biological weapons at the United Nations\footnote{https://www.washingtonpost.com/world/2022/03/11/un-council-ukraine-russia-chemical-weapons-zelensky/}, the Russian narrative about bio-weapon labs swelled with new stories of new-found ``documents'' proving the existence of ``biolaboratories'' created and financed by the United States in Ukraine, where experiments were conducted with samples of bat coronavirus.''
\vspace{-1mm}
\paragraph{Narrative Themes Persist Across Stories}
Moreover, we find  that the stories within narratives are typically characterized by 1-3 closely related themes, reinforcing the concept of these clusters as the foundational elements of narratives. The prevalence of multiple themes seems to stem from the interconnectedness of these themes. For instance, within the denazification narrative, messages containing the theme ``Ukrainian leaders depicted as Nazi sympathizers'' were significantly more likely to have the theme ``Ukrainian nationalism linked with neo-Nazism'' (with a Spearman correlation coefficient of 0.83 with a p-value $<$ 0.01). These types of correlations between similar themes were found in multiple narratives and, when coupled with the lack of correlations found among more disparate themes, appear to support the notion that narratives may act as frameworks
which may align otherwise disparate domains of knowledge \cite{corona-conspiracy}.
\vspace{-1mm}
\paragraph{Narrative Focuses May Correlate with Key Contributors}
We then explore associations between contributors' posting patterns and subsequent narrative shifts in online communities. Our analysis begins with Granger causality tests to identify whether the timing of an author’s posts can forecast changes in engagement with the narrative. We ensure that the author's own posts are excluded from later analyses to prevent bias. This analysis highlights statistically significant predictive relationships (p-value $<$ 0.01), but it is important to note that these do not confirm causal relationships.

To further examine these relationships, we use Spearman's rank correlation coefficient, appropriate for assessing non-linear associations. Results are detailed in Tables \ref{tab:narrative_associations_general} and \ref{tab:narrative_theme_associations}, focusing on correlations that exceed a threshold of 0.3 and are significant at a p-value of less than 0.01. We find that the most notable associations for overall narrative shifts typically occur with a one-day lag, illustrating the rapid influence of specific contributions. However, the timing of peak associations varies more in the context of thematic engagement, with some authors demonstrating significant correlations several days later. Interestingly, those with high thematic correlations are not always among the most prolific posters, suggesting a complex interplay in how narratives evolve within the community.
\begin{table}[h!]
\centering
\small
\vspace{-1mm}
\caption{Narrative story associations across pro-Russian and pro-Ukrainian communities (p-value is $< 0.01$).}
\label{tab:narrative_associations_general}
\begin{tabular}{llcc}
\toprule
Narrative & Channel & Lag & Spearman \\
\midrule
\multicolumn{4}{c}{\textbf{Russian}}\\
\midrule
Denazification & boris\_rozhin & 1 & 0.461 \\ 
Biological Weapons & regnum\_na & 1 & 0.395 \\
Biological Weapons & regnum\_na & 3 & 0.318 \\
Bucha - Russia & SolovievLive & 2 & 0.416 \\
\addlinespace
\multicolumn{4}{c}{\textbf{Ukrainian}} \\
\midrule
Human Corridors & liganet & 1 & 0.423 \\
EU Accession & u\_now & 7 & 0.350 \\
\bottomrule
\end{tabular}
\end{table}
\vspace{-2mm} 

\begin{table*}[]
\centering
\setlength{\tabcolsep}{4pt} 
\small
\caption{Narrative theme associations across pro-Russian and pro-Ukrainian communities (p-value is $< 0.01$).}
\label{tab:narrative_theme_associations}
\begin{tabular}{lllcc}
\toprule
Narrative & Theme & Channel & Lag & Spearman \\
\midrule
\multicolumn{5}{c}{\textbf{Russian}} \\
\midrule
BioWeapons & Military-Bio Activities & SolovievLive & 1 & 0.439 \\
BioWeapons & Human Right Violations & SolovievLive & 3 & 0.453 \\
BioWeapons & Involvement of US & SolovievLive & 4 & 0.351 \\
Denazification & Neo-Nazi Policies Effects & boris\_rozhin & 1 & 0.436 \\
Bucha - Russia & Bucha Conspiracy  & SolovievLive & 4 & 0.320 \\
\addlinespace

\multicolumn{5}{c}{\textbf{Ukrainian}} \\
\midrule
Human Corridors & Humanitarian Aid & znua\_live & 4 & 0.513 \\
Human Corridors & International Cooperation & znua\_live & 4 & 0.506 \\
Bucha - Ukrainian & Humanitarian Crisis & u\_now & 2 & 0.303 \\
\bottomrule
\end{tabular}
\end{table*}

\begin{table*}[t!]
\centering
\vspace{-1mm} 
\caption{Pro-Ukrainian information narrative and theme analysis.}
\vspace{-1mm} 
\label{tab:uk_narratives}
\begin{tabular}{p{4cm}p{9cm}p{3cm}}
\toprule
Narrative & Narrative Themes Discovered & Top Contributing Authors \newline (Contribution / Std Devs Above Median)  \\
\midrule
Russian troops committed a massacre in Bucha. & Russian Atrocities, Ukrainian Resistance, International Diplomacy, Humanitarian Crisis, Information Warfare, War Crimes Accountability, Economic Impact, Refugees, Bucha Comparisons. & kyiv\_n (4.4\% / 5.8)\newline u\_now (3.0\% / 3.8) \\
\midrule
Ukraine has a unique European identity and belongs in the European Union. & Ukraine's EU Aspiration, Accession Challenges, International Support, EU Bilateral Relations, EU Integration Progress.& verkhovnaradaukrainy (7.8\% / 8.52) \newline u\_now (3.7\% / 3.9) \\
\midrule
Russia is engaging in chemical warfare. & Chemical Weapon Concerns, Reports of Usage, Preparation, Environmental Impact, Anti-Mite Treatment.& ukraina24tv (4.04\% / 4.6) \newline spravdi (3.1\% / 3.3)  \\
\midrule
Russia is sabotaging Ukrainian humanitarian corridors.& Humanitarian Aid, Evacuations, Corridor Blockades, War Crimes Probes, International Cooperation, Government Actions, International Support.& znua\_live (4.1\% / 3.3) \newline OP\_UA (4.0\% / 3.2)  \\
\bottomrule
\end{tabular}
\end{table*}

\vspace{-1mm} 
\begin{table*}[t!]
\centering
\vspace{-1mm} 
\caption{Pro-Russian information narrative and theme analysis.}
\vspace{-1mm} 
\label{tab:ru_narratives}
\begin{tabular}{p{4cm}p{9cm}p{3cm}}
\toprule
Narrative & Narrative Themes Discovered & Top Contributing Authors \newline (Contribution / Std Devs Above Median) \\
\midrule
There are Biological Weapons Laboratories in Ukraine. & Ukrainian Bio-Weapons Labs, Military Activities, Global Involvement, Pathogens, Human Rights Issues, Safety Concerns, Strategic Locations, Information Warfare, Health Threats, Legal Action. & regnum\_na (5.8\% / 6.2)\newline SolovievLive (4.5\% / 4.8)\\
\midrule
Ukraine must be denazified. &  Ukrainian Neo-Nazism, WWII History Revision, Russian Anti-Nazi Stance, Western Support, Leadership Sympathies, Conflict Symbols, Media Propagation, Patriotism, Neo-Nazi Impact, War Crimes, Civilian Effects & SolovievLive (3.6\% / 6.0)\newline rus\_demiurge (3.2\% / 5.3) \\
\midrule
There was no ``Bucha Massacre'' committed by Russian troops. &      Bucha Event Doubts, Media Manipulation, International Report Challenges, Alternative Narratives, Ukraine's Role Critique, Evidence Credibility, International Response & SolovievLive (4.9\% / 5.5) \newline borish\_rozhin (3.7\% / 4.2)\\
\midrule
NATO poses a threat to Russia.  &          NATO Expansion, Military Provocations, Western Aggression, Conflict Escalation, Hypocrisy, Indirect Warfare, Geopolitical Impact, Global Security Threat, Russia's Defense. & rus\_demiurge (4.2\% / 5.81) \newline SolovievLive (3.1\% / 4.2) \\
\midrule
\end{tabular}
\vspace{-1mm}
\end{table*}

\paragraph{Narratives Adapt in Quantity and Makeup} Narratives appear to adapt to external influences, showing changes in both post volume and thematic content. This adaptation is evident in how narratives respond to events, incorporating both the frequency of posts and their evolving themes.

The Bucha Massacre serves as a distinct example of this phenomenon. The Ukrainian community's narrative was characterized by active reporting on the event, focusing on civilian harm and military defense, as shown in the increase of posts during this time. Conversely, the Russian community initially minimized discussion, later shifting to claims of "provocation" and misinformation as the event gained international attention. This divergence in response and strategy between communities is starkly illustrated in the aftermath of Bucha, emphasizing the narratives' dynamic nature.

\begin{figure}[h!]
\centering
\small
\begin{tikzpicture}
\begin{axis}[
    xlabel={Date},
    legend style={at={(0.5,-.5)},anchor=north},
    ylabel={Score},
    title={Scores Over Time for Different Categories},
    date coordinates in=x, 
    xticklabel={\year-\month-\day}, 
    xticklabel style={rotate=50, anchor=east}, %
    xlabel style={
        yshift=-7.5pt, 
    },
    ylabel=Proportion of All Posts,
    xlabel=Date (2022),
    title=Top 4 Themes in Russian Denazification Narrative,
    grid=both, 
    width=\linewidth, 
    height=4.5cm, 
    xtick={2022-02-20, 2022-03-12, 2022-04-01, 2022-04-21, 2022-05-11, 2022-05-29},
    xticklabels={Feb 20, Mar 12, Apr 1, Apr 21, May 11, May 29},
    xmax=2022-05-29, %
    xmin=2022-02-20, %
    ]


\definecolor{frenchblue}{rgb}{0.0, 0.45, 0.73}
\definecolor{frenchlilac}{rgb}{0.53, 0.38, 0.56}
\definecolor{frenchrose}{rgb}{0.96, 0.29, 0.54}
\definecolor{lava}{rgb}{0.81, 0.06, 0.13}
\definecolor{lavenderindigo}{rgb}{0.58, 0.34, 0.92}
\definecolor{maize}{rgb}{0.98, 0.93, 0.37}
\definecolor{green(munsell)}{rgb}{0.0, 0.66, 0.47}
\definecolor{internationalorange}{rgb}{1.0, 0.31, 0.0}
\definecolor{officegreen}{rgb}{0.0, 0.5, 0.0}
\definecolor{mustard}{rgb}{1.0, 0.86, 0.35}

\addplot+[
  mark=none,
  mark options={solid},
  color=frenchblue
  ]
  table [x=date, y=Accusations of war crimes by Ukrainian forces, col sep=comma] {LaTeX/nazis_plottable.csv};
\addlegendentry{Western Support for Nazism}


\addplot+[
  mark=none,  
  mark options={solid},  
  color=lavenderindigo
  ]
  table [x=date, y=Propagation of neo-Nazi themes in media, col sep=comma] {LaTeX/nazis_plottable.csv};
\addlegendentry{Russian Invasion as Fight Against Nazism}
\definecolor{electricviolet}{rgb}{0.56, 0.0, 1.0}

\addplot+[
solid,
  mark=none,  
  mark options={solid},  
  color=lava
  ]
  table [x=date, y=Effects of neo-Nazi policies on civilians, col sep=comma] {LaTeX/nazis_plottable.csv};
\addlegendentry{Effects of neo-Nazi policies on civilians}

\addplot+[
    solid,
  mark=none,  
  mark options={solid},  
  color=officegreen
  ]
  table [x=date, y=Ukrainian nationalism linked with neo-Nazism, col sep=comma] {LaTeX/nazis_plottable.csv};
\addlegendentry{Ukrainian nationalism linked with neo-Nazism}

\end{axis}

\end{tikzpicture}
\caption{Themes discovered in narratives about Denazifying Ukraine.}
\label{fig:ru_denazify}
\vspace{-0.3cm}
\vspace{-2mm} 
\end{figure}
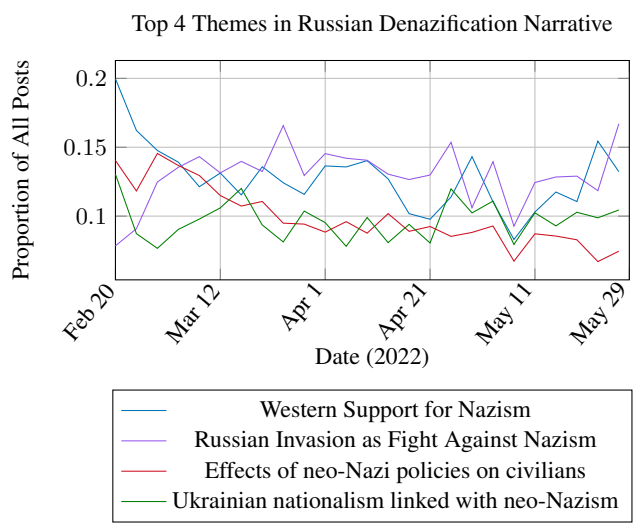

The evolution of narrative themes further demonstrates their responsiveness. Initially, Russia's narrative centered on "denazification," highlighting issues like language suppression and cultural genocide, as referenced in early statements\footnote{https://en.wikipedia.org/wiki/Stepan\_Bandera}. However, as the conflict progressed, the narrative pivoted to patriotic themes, celebrating Russia's actions against perceived Nazism, as depicted in Figure \ref{fig:ru_denazify}. This shift from focusing on the "Nazification of Ukraine" to lauding "liberation" efforts showcases the narratives' capacity to adapt to evolving contexts and external perceptions.



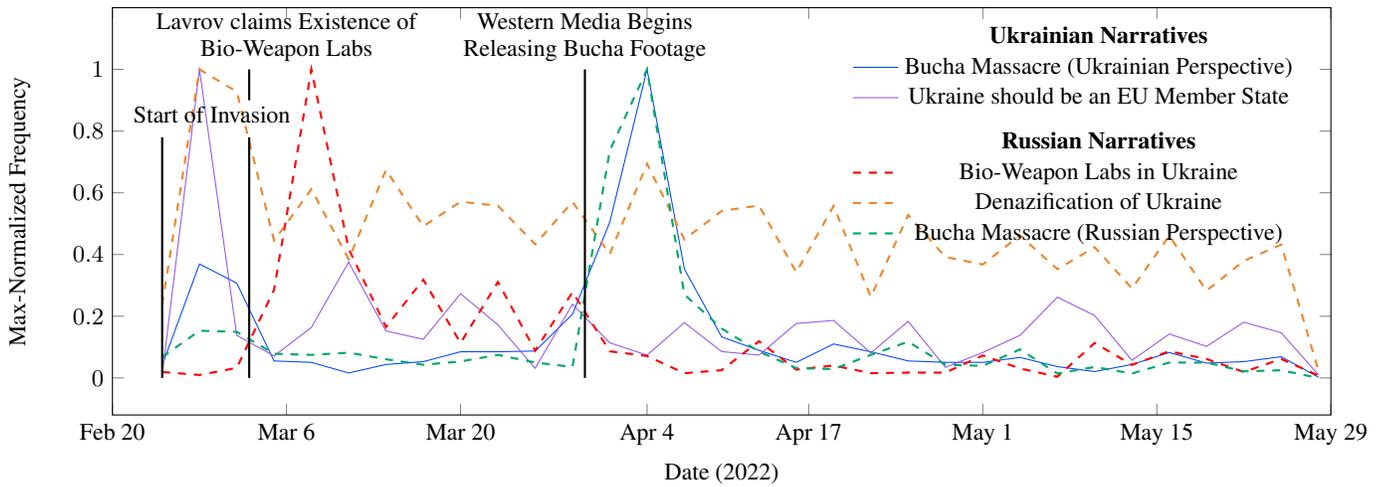
\begin{figure*}[t]
\small
\centering

\begin{tikzpicture}
\begin{axis}[
    xlabel={Date},
    date coordinates in=x, 
    xticklabel style={rotate=0, 
    anchor=north}, %
    ylabel=Max-Normalized Frequency,
    xlabel=Date (2022),
    width=\linewidth, 
    height=7cm, 
        legend style={fill=none,
        draw=none, 
    },
    xtick={2022-02-20, 2022-03-06, 2022-03-20, 2022-04-04, 2022-04-17, 2022-05-01, 2022-05-15, 2022-05-29},
    xticklabels={Feb 20, Mar 6, Mar 20, Apr 4, Apr 17, May 1, May 15, May 29},
    ytick={0, 0.2, 0.4, 0.6, 0.8, 1},
    xmax=2022-05-29, %
    xmin=2022-02-20, %
    ymax=1.2, 
    ]
    
    Russian Military Actions and Atrocities
Comparative References to Bucha
\addlegendimage{empty legend}
\addlegendentry{\textbf{Ukrainian Narratives}}

\definecolor{airforceblue}{rgb}{0.36, 0.54, 0.66}
\definecolor{blizzardblue}{rgb}{0.67, 0.9, 0.93}
\definecolor{blue(ryb)}{rgb}{0.01, 0.28, 1.0}
\definecolor{bostonuniversityred}{rgb}{0.8, 0.0, 0.0}
\addplot+[
  mark=none,
  mark options={solid},
  color=blue(ryb)
  ]
  table [x=date, y=Bucha Massacre (Ukrainian Perspective), col sep=comma] {LaTeX/narratives_plot.csv};
\addlegendentry{Bucha Massacre (Ukrainian Perspective)}
\definecolor{electriccrimson}{rgb}{1.0, 0.0, 0.25}

\definecolor{capri}{rgb}{0.0, 0.75, 1.0}
\definecolor{languidlavender}{rgb}{0.84, 0.79, 0.87}
\definecolor{lavenderindigo}{rgb}{0.58, 0.34, 0.92}
\addplot+[
solid,
  mark=none,
  mark options={solid},
    color=lavenderindigo
  ]
  table [x=date, y=Ukraine should be an EU Member State, col sep=comma] {LaTeX/narratives_plot.csv};
\addlegendentry{Ukraine should be an EU Member State}

\addlegendimage{empty legend}
\addlegendentry{}

\addlegendimage{empty legend}
\addlegendentry{\textbf{Russian Narratives}}
\addplot+[
dashed,
thick,
  mark=none,
  mark options={solid},
  color=red
  ]
  table [x=date, y=Bio-weapon Labs in Ukraine, col sep=comma] {LaTeX/narratives_plot.csv};

\addlegendentry{Bio-Weapon Labs in Ukraine}

\definecolor{byzantium}{rgb}{0.44, 0.16, 0.39}
\definecolor{cadmiumorange}{rgb}{0.93, 0.53, 0.18}
\addplot+[
dashed,
thick,
  mark=none,
  mark options={solid},
  color=cadmiumorange
  ]
  table [x=date, y=Denazification of Ukraine, col sep=comma] {LaTeX/narratives_plot.csv};
\addlegendentry{Denazification of Ukraine}

\definecolor{dogwoodrose}{rgb}{0.84, 0.09, 0.41}
\definecolor{electricgreen}{rgb}{0.0, 1.0, 0.0}
\definecolor{jade}{rgb}{0.0, 0.66, 0.42}

\addplot+[
dashed,
thick,
  mark=none,  
  mark options={solid},  
    color=jade
  ]
  table [x=date, y=Bucha Massacre (Russian Perspective), col sep=comma] {LaTeX/narratives_plot.csv};
\addlegendentry{Bucha Massacre (Russian Perspective)}


\node[above, align=center] at (axis cs:2022-03-30,1) {Western Media Begins\\ Releasing Bucha Footage};

\draw [thick] (axis cs:2022-03-30,0) -- (axis cs:2022-03-30,1); 

\node[above] at (axis cs:2022-02-28,.8) {Start of Invasion};
\draw [thick] (axis cs:2022-02-24,.78) -- (axis cs:2022-02-24, 0); 

\node[above, align=center] at (axis cs:2022-03-06,1) {Lavrov claims Existence of\\Bio-Weapon Labs};
 
\draw [thick] (axis cs:2022-03-03,0) -- (axis cs:2022-03-03,.78); 
\draw [thick] (axis cs:2022-03-03,.9) -- (axis cs:2022-03-03,1); 

\end{axis}

\end{tikzpicture}
\small
\vspace{-4mm}
\caption{Information narrative evolution shown using max-normalized frequency of posts in key narratives within Ukrainian and Russian communities over time (summed over 3-day time-periods for clarity), where max-normalization adjusts frequency counts relative to each narrative's peak activity for comparative clarity.}
\label{fig:overarching_narratives}
\vspace{-1mm}
\end{figure*}

\begin{figure*} 
\centering
\includegraphics[width=1.1\textwidth]{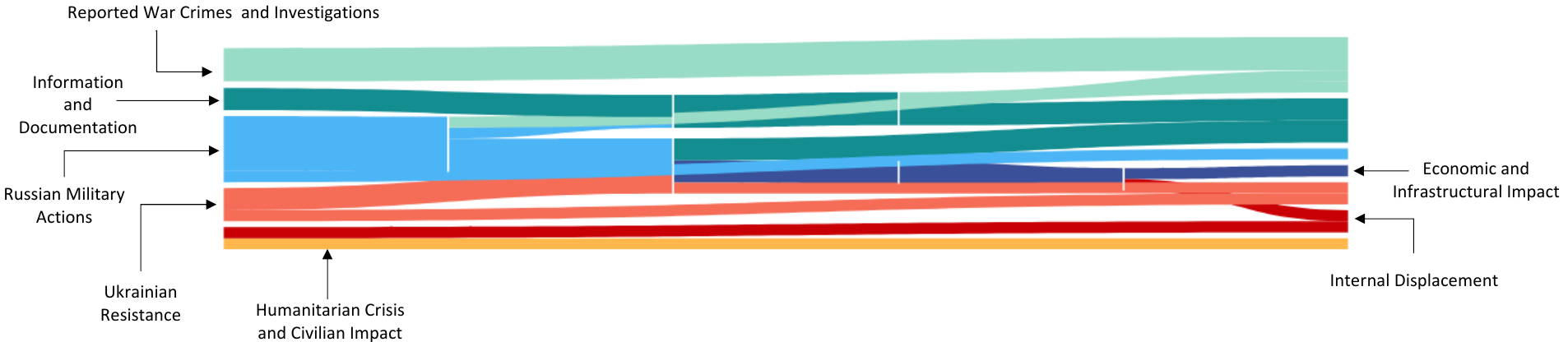} 
\caption{Information narrative evolution of dominant theme distribution of story clusters focused on Russian atrocities. To make this we look at the top 2 dominant themes in top 5 most populated story clusters; then, we aggregate this at a monthly timeframe (thus, having 4 points for potential pivoting) to understand general flow of these themes.} 
\vspace{-2mm} 
\label{fig:theme_flow_diagram}
\vspace{-2mm} 
\end{figure*}

\paragraph{Russian and Ukrainian Communities' Perceptions Differ Strikingly: Bucha Massacre Case Study}
The diverging narratives around the Bucha Massacre, as depicted in Figures \ref{fig:bucha_narratives} and \ref{fig:overarching_narratives}, highlight the distinct ways each community perceives the events. From the Ukrainian perspective, the narrative starts unfolding in real-time, with a focus on documenting the atrocities. This emphasis on documentation persists and intensifies over time, becoming crucial to counter the Russian narrative's accusations of fabrication and misinformation (this general shift from on-the-ground coverage to documentation is further illustrated in Figure \ref{fig:theme_flow_diagram}). Additionally, this framework of understanding becomes a lens through which the Ukrainian community views and discusses other similar incidents, increasingly comparing other events to the massacre with statements like ``the second Bucha is now taking place in the Kherson region,'' indicating a broader application of the Bucha narrative to comprehend subsequent events. Contrastingly, the Russian narrative on the Bucha Massacre, emerging predominantly post-Western media coverage, consistently features claims of propaganda and Ukrainian provocation. This emphasis on efforts to ``expose the fake about Bucha'' and challenge the ``demonization of Russia'' by Western media intertwines with a persistent theme of Russian victimhood, which we note across other narratives discussion by Russian communities.

\vspace{-2mm} 
\section{Discussion}

Our deeper exploration into the narratives discussed by Russian and Ukrainian communities on Telegram during the early stages of the war sheds light on the dynamic and contrasting ways these groups perceive and respond to unfolding events. This analysis has revealed how quickly online communities respond to external events, highlighting the inherently fluid nature of digital narratives. For example, the distinct focuses on military actions, political movements, and humanitarian efforts between Russian and Ukrainian narratives not only showcase divergent priorities but also underscore the significant impact of narrative framing on public engagement and perception. Additionally, the contrasting reactions to events like the Bucha Massacre, coupled with the varied emphasis on themes such as "denazification" and "bioweapon labs," reveal a complex interplay of information, misinformation, and narrative communication and dissemination strategies. These dynamics are consistent with previous research on these topics \cite{golovchenko2018state}, reinforcing the need for nuanced, dynamic approaches in narrative analysis. Such approaches must be capable of accommodating rapid shifts in online discourse and addressing the complex, evolving nature of narratives that propagate online. 

Moreover, our study's exploration of narrative dynamics within these communities highlights the pivotal role certain key contributors play in either advancing or trailing the trends in narrative development. While these findings do not explicitly establish causation, they suggest potential alignment with theories that evaluate the influence of elites in shaping opinions, as discussed in previous studies \cite{chong2007theory}.

Finally, our approach highlights the limitations of traditional narrative analysis methods, which often struggle to capture the dynamic and rapidly evolving nature of digital narratives as well as the machinations of their inner components. In contrast, our study introduces a dynamic clustering model specifically designed to analyze textual data over time at scale, effectively adapting to and reflecting the fluidity of online narratives. Our approach works to overcome the limitations of other models and track the genesis, evolution, and impact of narratives, offering a methodological advancement over static methods that fail to keep pace with the ever-changing digital discourse (see our Related Work section for more details on SOTA approaches). 


\vspace{-2mm} 
\section{Ethical Considerations}
Our study utilized publicly accessible data from influential figures discussing the Russia-Ukraine war on Telegram, and we approached the sensitive subject matter with respect and objectivity. While we aim to advance computational narrative analysis by releasing our \texttt{OnlineAgglomerative} class and Jupyter notebooks, we recognize the potential for misuse and emphasize the importance of responsible application, prioritizing privacy and well-being. Our findings are specific to the studied online communities and should not be generalized. We have taken steps to prevent misuse by limiting data to public posts, excluding personally identifiable information, and providing clear usage guidelines. Ethical considerations have been central throughout our research process, and we prioritize the responsible and ethical use of these tools and insights.


\vspace{-1mm} 
\section{Limitations}


First, our methodology's effectiveness depends on similarity metrics derived from text embeddings for clustering narratives. The performance of these metrics is tied to effectiveness of embedding models, which may not perform well with contexts different from their training data or with lengthy texts, potentially limiting clustering outcomes. We have made our approach modular to allow the integration of alternative embedding models  that could better capture the nuances of specific contexts and operational domains \cite{devlin2018bert, reimers2019sentencebert, gao2022simcse}. Additionally, our study focuses only uses the initial three months of the Russia-Ukraine war data for demonstration of the efficacy of the proposed approach, and do not intend to reflect the long-term evolution of narratives during the war. Extending the observation period could yield deeper insights into the dynamic processes of narrative evolution, enhancing our understanding of their life cycles.

Next, the study relies on a sample of Telegram channels and posts, which may not be representative of the entire population of users discussing the Russia-Ukraine conflict. The selection criteria for channels and the time period chosen could introduce sampling biases that limit the generalizability of the findings. Future work should consider expanding the sample size, diversifying the channels included, and exploring different sampling techniques to mitigate potential biases. Finally, the methodology focuses primarily on the text content of posts and lacks incorporation of other contextual factors such as images, videos, and links shared within the Telegram communities. These multimedia elements may provide additional insights into narrative dynamics and could influence the interpretation of the findings. Future research should explore methods for integrating and analyzing multimedia content alongside text data.


\vspace{-1mm} 
\section{Summary and Future Work}

Our study introduces a novel methodology for the dynamic analysis of online narratives, which proved effective when applied to discussions within Ukrainian and Russian Telegram communities during the early stages of the Russia-Ukraine war. The approach successfully captured and contrasted the evolution of narratives, adapting to rapid changes in online discourse. The case study demonstrated the method's ability to identify narrative strategies, themes, and key contributors, providing valuable insights into information flow dynamics and public perception. This research offers significant benefits for Operations in the Information Environment (OIE), empowering planners and analysts to swiftly adapt strategies, optimize resources, and effectively engage target audiences, ultimately enhancing decision-making, situational awareness, and communication strategies in complex information environments.

Future research should focus on expanding the methodology's application to diverse platforms and operational contexts, while addressing limitations through collaborative efforts between domain experts, data scientists, and practitioners. For example, integrating multimedia content analysis alongside text data will provide a more comprehensive understanding of narrative dynamics. Where appropriate, incorporating user demographic data integrated into deeper cognitive domain analysis of targeted audience e.g., emotions, reactions, perceptions, moral values etc. can offer insights into how narratives evolve and perceived by different groups. By addressing these limitations, and integrating emerging technologies, future work can enhance the accessibility, usability, and depth of dynamic narrative analysis, ultimately contributing to a more nuanced understanding of information flows, public opinion formation, and the interplay between online narratives and offline real-world events.

\vspace{-2mm} 
\bibliography{LaTeX/aaai24}

\end{document}